\documentclass[aps,twocolumn,showpacs]{revtex4}

\usepackage{graphicx}
\usepackage{amsmath}
\usepackage{amssymb}
\usepackage{amsbsy}
\usepackage{latexsym}
\usepackage{dcolumn}
\usepackage{bm}

\begin{document}

\title{An efficient finite element method applied to quantum billiard systems}

\author{Woo-Sik \surname{Son}}

\email{dawnmail@sogang.ac.kr}

\affiliation
{Acceleration Research Center for Quantum Chaos Applications,\\
Department of Physics, Sogang University, Seoul 121-742, Korea}

\author{Sunghwan \surname{Rim}}

\email{rim001@sogang.ac.kr}

\affiliation
{Acceleration Research Center for Quantum Chaos Applications,\\
Department of Physics, Sogang University, Seoul 121-742, Korea}

\author{Chil-Min \surname{Kim}}

\email{chmkim@sogang.ac.kr}

\affiliation
{Acceleration Research Center for Quantum Chaos Applications,\\
Department of Physics, Sogang University, Seoul 121-742, Korea}

\begin{abstract}
An efficient finite element method (FEM) for calculating eigenvalues
and eigenfunctions of quantum billiard systems is presented. We
consider the FEM based on triangular $C_1$ continuity quartic
interpolation. Various shapes of quantum billiards including an
integrable unit circle are treated. The numerical results show that
the applied method provides accurate set of eigenvalues exceeding a
thousand levels for any shape of quantum billiards on a personal
computer. Comparison with the results from the FEM based on
well-known $C_0$ continuity quadratic interpolation proves the
efficiency of the method.
\end{abstract}


\maketitle

\section{INTRODUCTION}

There has been much interest in characterizing the quantum
manifestation of classically chaotic systems
\cite{Gutzwiller,Stockmann}, since McDonald and Kaufman's pioneering
investigation on the statistical characteristics of eigenvalues and
eigenfunctions \cite{McDonald}. The quantum billiard, which is
represented by the two dimensional stationary Schr\"{o}dinger
equation of free particle with satisfying the Dirichlet boundary
condition, is an intensively studied model system in the field of
quantum chaos due to its simplicity. The integrability of
corresponding classical billiard depends solely upon the geometry of
boundary. The quantum billiard can be also expressed by the scalar
Helmholtz equation, for example, which describes the electromagnetic
field inside a flat microwave resonator. In that context, microwave
experiments played the role of {\it analog} computation of
eigenstates in quantum billiards \cite{Stockmann1,Sridhar}.

There are several numerical methods, which have been dominantly
adopted by colleagues in this field, for calculating eigenvalues and
eigenfunctions of the quantum billiards such as the boundary
integral method (BIM) (reviewed in Ref. \cite{Backer}), the plane
wave decomposition method (PWDM) \cite{Heller,Baowen}, the scaling
method \cite{Vergini,Barnett}, and the conformal mapping method
\cite{Robnik,Berry,Prosen}. Let us briefly review the mentioned
methods. The BIM, which is a rigorously established method, reduces
the problem of two dimensional stationary Schr\"{o}dinger equation
to a one dimensional integral equation. In result, each root of the
Fredholm determinant constitutes the set of eigenvalues. In
practice, the determinant does not become zero due to the
discretization error and the BIM approximates the minima of the
lowest singular values to the eigenvalues. Though many successful
applications, the BIM has some shortcomings. One of them is that the
calculated results can include additional, i.e., spurious solutions
which correspond to roots of outside scattering problem with Neumann
boundary condition \cite{Tasaki}. For non-convex geometric billiard,
the detection of spurious solutions is not a simple numerical task
\cite{Backer} and the BIM failed in the isospectral drum introduced
by Gordon, Webb, and Wolpert \cite{Gordon} due to its strong
non-convexity \cite{Baowen1}. The feasibility of missing eigenvalues
is an another weakness of the BIM. For higher lying eigenvalues, the
spectra become more denser and the correct detection of minima is a
serious problem.

The PWDM, which has been introduced by Heller \cite{Heller}, is a
rather heuristic method in the context of quantum chaos. It is
appropriate for computing high lying eigenstates but incongruent for
studying of spectral statistics, because only a few selected
eigenstates can be calculated with many intermediate missing. Also
the PWDM can fail in non-convex or multiply connected (e.g.,
containing a hole) billiards \cite{Gutkin}. In the literatures,
there has been a considerably efficient numerical method, that is,
the scaling method derived by Vergini and Saraceno \cite{Vergini}.
It represents the boundary norm as a function of energy by the use
of {\it scaling}. In result, the authors of Refs.
\cite{Vergini,Barnett} obtained all eigenvalues (without any
missing) within a narrow energy range, which lie close to a chosen
reference value, in a single computational step. The efficiency of
scaling method is obvious from that the BIM can locate a single
eigenvalue in a single computational step. For specific geometric
billiard for which the conformal mapping onto the unit disk is
sufficiently simple (e.g., so-called Robnik billiard
\cite{Robnik1}), the conformal mapping method derived by Robnik
\cite{Robnik} have provided accurate set of eigenvalues
\cite{Baowen1}. Recently, new approaches that combine each ideas of
above mentioned methods have been studied, for the BIM and the PWDM
\cite{Cohen} and for the BIM and the scaling method \cite{Veble}.
Concerning the scattering quantization method, an efficient
improvement has been carried out in Ref. \cite{Tureci}.

The finite element method (FEM) is one of the most widely accepted
numerical methods for partial differential equations in various
fields of science and engineering
\cite{Reddy,Ram-Mohan,Zienkiewicz}. Compared with previously
mentioned methods, the FEM has obvious advantages that it has almost
no limitation on the geometric complexity of billiard (see the
results in Ref. \cite{Menezes}) and provides in a single
computational step a set of all eigenvalues and eigenfunctions up to
maximal level allowed by memory allocation. However, the accurate
computation of high-lying eigenstates using the FEM is
conventionally more difficult than the case of other mentioned
methods, since the FEM discretizes not only the boundary but the
whole domain of billiard (it needs more memory storage). Thereby,
though its obvious advantages, the FEM has been apparently
overlooked in the field of quantum chaos. As far as we know, there
have been only a few studies \cite{Baez,Aguiar,Heuveline,Dietz,Menezes}
where the FEM is used for calculating the eigenstates of quantum
billiards. Among those studies, Heuveline showed an effective FEM
only requiring $O(N)$ memory allocation by using the p-finite
elements basis and the sparsity of matrices \cite{Heuveline,Dietz}.
Note that the FEM commonly gives rise to sparse matrices but usual
FEMs do not take advantage of the sparsity (it contains quite
difficult numerical tasks) and need $O(N^{2})$ memory storage, where
$N$ is the number of total nodes.

The aim of this paper is to show the validity of FEM as a numerical
method for calculating eigenvalues and eigenfunctions of quantum
billiards. For that purpose, we present an efficient FEM based on
the Hermite interpolation. In each element, the wave function is
interpolated by quartic polynomials involving nodal values of wave
function and its first derivatives, namely, the adopted
interpolation basis admits the $C_1$ continuity. By applying the
method, we calculate the eigenvalue spectra of unit disk
(integrable), the Robnik billiard (convex geometric chaotic), and
the spiral-shaped billiard (non-convex geometric chaotic). We show
that the method provides accurate set of eigenvalues exceeding a
thousand levels for any shape of quantum billiards on a personal
computer. Comparison with the results from the FEM based on
well-known $C_0$ continuity quadratic interpolation proves the
efficiency of the applied method. Note that, by virtue of the $C_1$
continuity, the method handles well problem that treats values of
first derivatives of wave function at the boundary such as the
Neumann boundary condition.

The rest of paper is organized as follows. In Sec. II, we outline
numerical procedures of the FEM based on $C_1$ continuity quartic
interpolation. The results of numerically calculated eigenvalues and
the analysis of spectral statistics for unit disk, the Robnik
billiard, and the spiral-shaped billiard are presented in Sec. III.
Conclusions are given in Sec. IV.

\section{Numerical Procedure}

The quantum billiard is governed by two-dimensional stationary
Schr\"{o}dinger equation of free particle

\begin{eqnarray}\label{eq:1}
-\nabla^2\psi(\vec{r})=E\psi(\vec{r}), & \textrm{for} & \vec{r}\in\Omega
\end{eqnarray}

\noindent
with satisfying the Dirichlet boundary condition
$\psi(\vec{r})=0$ at the boundary of domain $\partial\Omega$.
Note that we use the natural units $\hbar=2m=1$.

The first step of applying FEM is that the domain of billiard
$\Omega$ is discretized into finite elements, i.e., mesh generation.
The shape of element and the number of nodes in each element are
determined according to the type of interpolation polynomials (i.e.,
shape functions). Here we consider the FEM based on the triangular
$C_1$ continuity quartic interpolation, which has been derived by
Specht \cite{Specht} and known that passes all patch test, i.e., a
condition for assessing FEM convergence for arbitrary mesh
configurations (see Chapter 11 of Ref. \cite{Zienkiewicz}).
Accordingly the domain of billiard is discretized into triangular
elements. In each element, there exist three nodes that locate at
vertices of triangle and each node has three degrees of freedom
correspond to wave function and its first derivatives ($\psi$,
$\partial\psi/\partial{x}$, and $\partial\psi/\partial{y}$). Thereby
each element has actually nine nodes. In each element the unknown
function, i.e., wave function $\psi(x,y)$ is represented as a linear
combination of shape functions multiplied by as-yet-unknown nodal
values of wave function and its first derivatives. The shape
function is defined only over a given element and has zero value at
outside of it. An explicit representation of shape functions will be
postponed for a while. The numerical procedure of FEM requires that
each node has three indices; local index $il$, element index $ie$,
and global index $i$. The mesh generation completes the mapping from
local and element index ($il_{th}$ node of $ie_{th}$ element) to
global index ($i_{th}$ global node).

On numerical calculations in this paper, we use two considerable and
freely available mesh generators; the DistMesh \cite{Persson} and
the Triangle \cite{Shewchuk}. The DistMesh is a Matlab based mesh
generator that finds node locations settled down a equilibrium state
in a truss structure. A geometry of domain is represented by the
{\it signed distance function} from node to closest boundary
$\partial\Omega$, negative inside the domain. It generates high
quality meshes, i.e., almost equilateral triangles but can be faced
with difficulty for complex geometric boundary. The Triangle is a
robust Delaunay refinement code. The user-supplied data, which
contains the information of nodes placing on the boundary, are
employed for specifying the domain of billiard. The quality of
meshes is controlled by the constraint of minimum angle (up to $34$
degree) and maximum size of triangle. The Triangle has almost no
limitation on the complexity of geometry.

Now we can take the next step. Eq. (1)
can be obtained from the condition that the action

\begin{equation}\label{eq:2}
A=\int{ds}\big{(}\nabla\psi^{*}(\vec{r})\cdot\nabla\psi(\vec{r})
-E\psi^{*}(\vec{r})\psi(\vec{r})\big{)}
\end{equation}

\noindent is minimized with respect to variation of
$\psi^{*}(\vec{r})$. Here $\psi(\vec{r})$ and $\psi^{*}(\vec{r})$
are considered to be two independent variables. Then, the action
integral of Eq. (2) is discretized into integrations over each
element as

\begin{equation}\label{eq:3}
A=\sum_{ie}^{ne}A^{(ie)}
\end{equation}

\noindent where $ne$ is the number of total elements. In Cartesian
coordinates, the discretized action integral $A^{(ie)}$ is
represented by

\begin{figure}[t]
\includegraphics{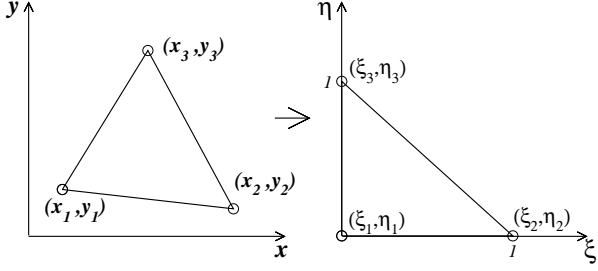}
\caption{A triangular real element and its mapping onto the parent
element, i.e, a right isosceles triangle.} \label{fig:fig1.eps}
\end{figure}

\begin{eqnarray}\label{eq:4}
A^{(ie)}&=&
\iint_{ie_{th}}dxdy\,\big( \frac{\partial\psi^{*}}{\partial x}\frac{\partial\psi}{\partial x}
+\frac{\partial\psi^{*}}{\partial y}\frac{\partial\psi}{\partial y}\big) {} \nonumber \\
&&{}-E\iint_{ie_{th}}dxdy\,\big(\psi^{*}\psi \big)
\end{eqnarray}

\noindent
where $\psi=\psi(x,y)$ and $\psi^{*}=\psi^{*}(x,y)$. For simple computation of $A^{(ie)}$,
it is advisable that the integral domain of each element is
transformed into a regularized domain (called parent element) as

\begin{equation}\label{eq:5}
A^{(ie)}=\iint_{ie_{th}}dxdy\,f(x,y)=\iint_{parent}d\xi d\eta\,g(\xi,\eta).
\end{equation}

\noindent Figure 1 shows a transformation of triangular real element
into the parent element. The quantities in Eq. (4) are altered into
$(\xi,\eta)$ notations as the followings. The wave function
$\psi(\xi,\eta)$ is interpolated by the $C_1$ continuity quartic
shape functions $\{H_{il}(\xi,\eta)\}$ multiplied by as-yet-unknown
nodal values of wave function and its first derivatives
$\{\tilde{\psi}_{il}\}$ as

\begin{eqnarray}\label{eq:6}
&&\psi(\xi,\eta)=\sum_{il}^{9}\tilde{\psi}_{il}H_{il}(\xi,\eta), {} \\
&&{}\tilde{\boldsymbol{\psi}}=
\{\psi_{1},\frac{\partial\psi_{1}}{\partial{x}},\frac{\partial\psi_{1}}{\partial{y}},
\psi_{2},\frac{\partial\psi_{2}}{\partial{x}},\frac{\partial\psi_{2}}{\partial{y}},
\psi_{3},\frac{\partial\psi_{3}}{\partial{x}},\frac{\partial\psi_{3}}{\partial{y}}\}. \nonumber
\end{eqnarray}

\noindent For simple representation, we use a cyclic property of the
applied shape functions and introduce the following notation
$\tilde{\boldsymbol{H}}_{a}=\left\{
H_{3a-2}(\xi,\eta),H_{3a-1}(\xi,\eta),H_{3a}(\xi,\eta) \right\}$ for
$a=1,2,3$. Then, the shape functions are represented by

\begin{eqnarray}\label{eq:7}
\tilde{\boldsymbol{H}}^{T}_{a}=\left\{ \begin{array}{c}
P_{a}-P_{a+3}+P_{c+3}+2(P_{a+6}-P_{c+6}) \\
m_{b}(P_{c+6}-P_{c+3})+m_{c}P_{a+6} \\
-n_{b}(P_{c+6}-P_{c+3})-n_{c}P_{a+6} \end{array} \right\}
\end{eqnarray}

\noindent
where $m_{a}=x_{c}-x_{b}$, $n_{a}=y_{b}-y_{c}$, and $a$, $b$, $c$ are
the cyclic permutations of 1, 2, 3.
The nine polynomials $\{P_{i}(\xi,\eta)\}$ in Eq. (7) are expressed as

\begin{eqnarray}\label{eq:8}
&&\boldsymbol{P}=
\Big\{ 1-\xi-\eta,\,\xi,\,\eta,\,(1-\xi-\eta)\xi,\,\xi\eta,\,\eta(1-\xi-\eta),{} \nonumber \\
&&{}\,\,\,(1-\xi-\eta)^{2}\xi+\frac{1}{2}(1-\xi-\eta)\xi\eta\cdot{} \nonumber \\
&&{}\,\,\,\big(3(1-\mu_{3})(1-\xi-\eta)-(1+3\mu_{3})\xi+(1+3\mu_{3})\eta\big),{} \nonumber \\
&&{}\,\,\,\xi^{2}\eta+\frac{1}{2}(1-\xi-\eta)\xi\eta\cdot{} \\
&&{}\,\,\,\big(3(1-\mu_{1})\xi-(1+3\mu_{1})\eta+(1+3\mu_{1})(1-\xi-\eta)\big),{} \nonumber \\
&&{}\,\,\,\eta^{2}(1-\xi-\eta)+\frac{1}{2}(1-\xi-\eta)\xi\eta\cdot{} \nonumber \\
&&{}\,\,\,\big(3(1-\mu_{2})\eta-(1+3\mu_{2})(1-\xi-\eta)+(1+3\mu_{2})\xi\big)\Big\} \nonumber
\end{eqnarray}

\noindent where $\mu_{a}=(l_{c}-l_{b})/l_{a}$ and
$l_{a}=(x_{b}-x_{c})^{2}+(y_{b}-y_{c})^{2}$ for the cyclic
permutation of $a$, $b$, $c$. The coordinate transformation between
real and parent elements is given by the following $C_0$ continuity
linear shape functions

\begin{equation}\label{eq:9}
x=\sum_{k}^{3}x_{k}^{(ie)}N_{k}(\xi,\eta),\,\, y=\sum_{k}^{3} y_{k}^{(ie)}N_{k}(\xi,\eta).
\end{equation}

\noindent where $N_{1}(\xi,\eta)=1-\xi-\eta$, $N_{2}(\xi,\eta)=\xi$,
and $N_{3}(\xi,\eta)=\eta$. Then, the infinitesimal surface $dxdy$
is represented by

\begin{equation}\label{eq:10}
dxdy=\left(\frac{\partial x}{\partial\xi}\frac{\partial y}{\partial\eta}
-\frac{\partial x}{\partial\eta}\frac{\partial y}{\partial\xi}\right)d\xi d\eta.
\end{equation}

\noindent The partial derivatives of wave function are transformed
into $(\xi,\eta)$ notations as follows:

\begin{eqnarray}\label{eq:11}
&&\frac{\partial\psi(x,y)}{\partial x}=
\frac{\partial}{\partial x}\left(\sum_{il}^{9}\tilde{\psi}_{il}^{(ie)}H_{il}(\xi,\eta)\right){}  \\
&&{}=\sum_{il}^{9}\tilde{\psi}_{il}^{(ie)}\left(\frac{\partial H_{il}(\xi,\eta)}{\partial\xi}
\frac{\partial\xi}{\partial x}+\frac{\partial H_{il}(\xi,\eta)}
{\partial\eta}\frac{\partial\eta}{\partial x}\right),{} \nonumber \\
&&{}\frac{\partial\psi(x,y)}{\partial y}=
\frac{\partial}{\partial y}\left(\sum_{il}^{9}\tilde{\psi}_{il}^{(ie)}H_{il}(\xi,\eta)\right){} \nonumber \\
&&{}=\sum_{il}^{9}\tilde{\psi}_{il}^{(ie)}\left(\frac{\partial
H_{il}(\xi,\eta)}{\partial\xi} \frac{\partial\xi}{\partial
y}+\frac{\partial H_{il}(\xi,\eta)}
{\partial\eta}\frac{\partial\eta}{\partial y}\right) \nonumber
\end{eqnarray}

By applying Eqs. (6)-(11), the discretized action integral $A^{(ie)}$ is represented by

\begin{eqnarray}\label{eq:12}
&&A^{(ie)}=\sum_{il,jl}^{9}\tilde{\psi}_{il}^{(ie)*}\left(\int_{0}^{1}d\eta\int_{0}^{1-\eta}d\xi\,
f(\xi,\eta)_{il,jl}\right)\tilde{\psi}_{jl}^{(ie)} {} \nonumber \\
&&{}-E\,\sum_{il,jl}^{9}\tilde{\psi}_{il}^{(ie)*}\left(\int_{0}^{1}d\eta\int_{0}^{1-\eta}d\xi\,
g(\xi,\eta)_{il,jl}\right)\tilde{\psi}_{jl}^{(ie)}
\end{eqnarray}

\noindent where $f(\xi,\eta)$ and $g(\xi,\eta)$ are sixth and eighth
order polynomials, respectively. By using the optimized quadrature
rule over the triangle, which has been derived by Dunavant
\cite{Dunavant}, Eq. (12) can be exactly integrated as

\begin{equation}\label{eq:13}
\int_{0}^{1}d\eta\int_{0}^{1-\eta}d\xi\,f(\xi,\eta)=
\sum_{i}^{\mathcal{N}}f(\xi_{i},\eta_{i})\cdot\omega_{i}
\end{equation}

\noindent where $(\xi_{i},\eta_{i})$ is a quadrature point and
$\omega_{i}$ is a weight. $\mathcal{N}$ equals 12 and 16 for sixth
and eighth order polynomials, respectively (see the table in Ref.
\cite{Dunavant}). Then we obtain the following result

\begin{equation}\label{eq:14}
A^{(ie)}=\sum_{il,jl}^{9}\tilde{\psi}_{il}^{(ie)*}\cdot\left(L_{il,jl}^{(ie)}
-EM_{il,jl}^{(ie)}\right)\cdot\tilde{\psi}_{jl}^{(ie)}.
\end{equation}

Now we add up the discretized action integral of Eq. (14) according to Eq. (3).
Then the action is represented as

\begin{equation}\label{eq:15}
A=\sum_{i,j}^{N}\tilde{\psi}_{i}^{*}\cdot\left(L_{i,j}-EM_{i,j}\right)\cdot\tilde{\psi}_{j}
\end{equation}

\noindent where $L_{i,j}$ ($M_{i,j}$) is a summation of all
$L_{il,jl}^{(ie)}$ ($M_{il,jl}^{(ie)}$) for which satisfies that
$il_{th}$ and $jl_{th}$ nodes of $ie_{th}$ element are mapped into
$i_{th}$ and $j_{th}$ global nodes, respectively. $N$ is the number
of total global nodes. In our case, it is equal to three times the
number of {\it physical} global nodes, since each node has three
degrees of freedom.

The adaptation of Dirichlet boundary condition is implemented as
follows. If the $k_{th}$ global node places at the boundary of
billiard $\partial\Omega$, the nodal value of wave function
$\tilde{\psi}_{k}$ equals zero. It requires that the entries of
$k_{th}$ column of $L$ and $M$ matrices become zero and also the
entries of $k_{th}$ row set to zero since $\tilde{\psi}_{k}^{*}=0$.
In practice, this is achieved by dropping the $k_{th}$ row and
column. Then, the dimension of $L$ and $M$ matrices are reduced by
$N-N_{b}$ where $N_{b}$ is the number of global nodes located at
$\partial\Omega$.

Now we vary the action $A$ with respect to nodal values
$\tilde{\psi}^{*}$ and invoke the principle of least action. In
result, we obtain the discretized version of stationary
Schr\"{o}dinger equation in the form of generalized eigenvalue
problem

\begin{equation}\label{eq:16}
L\,\Psi=EM\,\Psi
\end{equation}

\noindent where $L$ and $M$ are $N-N_{b}$ dimensional real and
symmetric matrices and $\Psi$ is a column matrix, which consists of
nodal values $\tilde{\psi}$. We use the LAPACK routine DSPGV
\cite{LAPACK} for solving the generalized eigenvalue problem. Note
that this routine requires about $2(N-N_{b})^{2}$ memory storage.
Finally we obtain the set of $N-N_{b}$ eigenvalues $\{E_{i}\}$ and
eigenfunctions $\{\Psi_{i}\}$ where $\Psi^{T}_{i}=\left(
\tilde{\psi}_1\,\, \tilde{\psi}_2\,\, \cdots \,\, \tilde{\psi}_N
\right)$ through restoring zero nodal values at nodes on
$\partial\Omega$.

\section{Numerical Results}

\subsection{The unit circle billiard}

We firstly consider an integrable billiard whose boundary is given
by the unit circle for testing an efficiency of the FEM presented in
Sec. II. In this case, the eigenvalue spectra are exactly known and
given by the sorted set of $\mu^{2}_{jk}$ where $\mu_{jk}$ is the
$k_{th}$ root of $j_{th}$ Bessel function of the first kind with
considering the degenerate case on $j \neq 0$ for $j=0,1,2,\cdots$,
and $k\in\mathbf{N}$.

We investigate the relative error between exact and numerically
calculated eigenvalues

\begin{equation}\label{eq:17}
\mathcal{E}_{rel}(i)=\frac{\vert \tilde{E}_{i}-E_{i} \vert}{\tilde{E}_{i}}
\end{equation}

\noindent where $\{\tilde{E}_{i}\}$ is the set of exact eigenvalues.
For the results obtained from the FEM based on well-known triangular
$C_0$ continuity quadratic interpolation (see Refs.
\cite{Reddy,Ram-Mohan} for its shape function and isoparametric
transformation), we also compute the relative error
$\mathcal{E}_{rel}$. In Fig. 2(a), we plot the relative error
$\mathcal{E}_{rel}$ for both cases of interpolation basis. For the
$C_1$ continuity quartic interpolation, we calculate $21500$ (equals
to $N-N_{b}$) eigenvalues and the result of relative error is drawn
by a black line. For the $C_0$ continuity quadratic interpolation,
we obtain $21647$ (almost same number of the above) eigenvalues and
depict the relative error as a orange line. Note that all
calculations in this paper are performed on a personal computer
possessing 2.4 GHz Quad-Core CPU and 8 GByte memory. So the
available maximal number of level allowed by memory allocation is
limited about $22000$. In both interpolations, the relative error
$\mathcal{E}_{rel}$ increases as the number of level $n$ increases.
However, the method based on the $C_0$ continuity quadratic
interpolation seriously loses its accuracy after a few hundreds
levels. The results of Fig. 2(a) proves an efficiency of the FEM
based on the $C_1$ continuity quartic interpolation.

We consider the numerical result for which the relative error
$\mathcal{E}_{rel}$ is smaller than $3\times10^{-3}$ as accurate
eigenvalue. In Fig. 2(b), the relative error for the $C_1$
continuity quartic basis is depicted up to $1500$ level and the
result is smaller than $3\times10^{-3}$ in this range. Then we
obtain {\it accurate} set of 1500 eigenvalues for the unit circle
billiard with this criterion. The followings will show that above
conjecture is reasonable.

\begin{figure*}[t]
\includegraphics{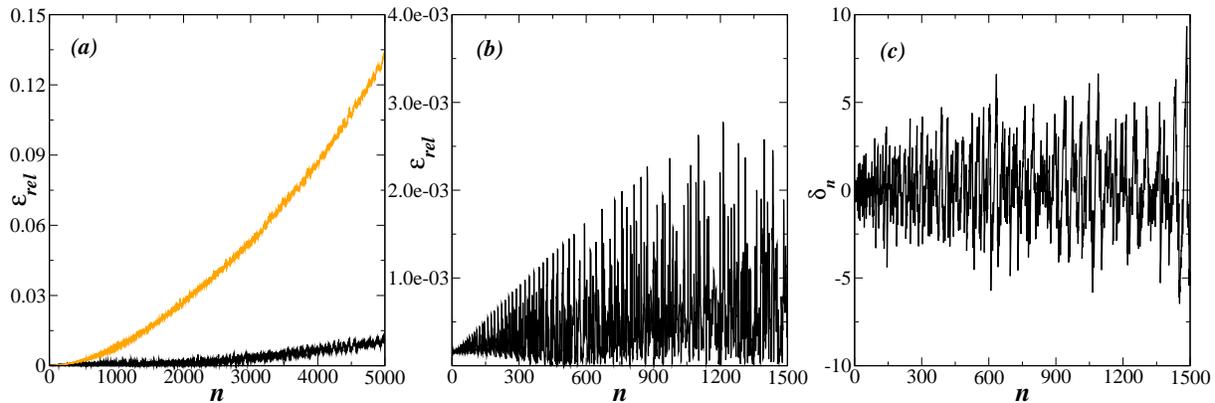}
\caption{(Color online) (a) The relative error $\mathcal{E}_{rel}$
for the $C_1$ continuity quartic interpolation (the $C_0$ continuity
quadratic interpolation) is drawn by a black (orange) line. (b) The
relative error $\mathcal{E}_{rel}$ up to $1500$ level for the $C_1$
continuity quartic basis. (c) The $\delta_{n}$ up to $1500$ level
for the $C_1$ continuity quartic basis.} \label{fig:fig2.eps}
\end{figure*}

Another method for testing the accuracy of obtained results is
checking out whether the eigenvalue spectra are complete without any
intermediate missing. For the system where the analytic eigenvalues
are not available, such method has no alternative. It can be
performed by investigating the spectral staircase function $N(E)$,
which counts the number of energy levels below $E$. The spectral
staircase function can be divided into a smooth and a fluctuating
part

\begin{equation}\label{eq:18}
N(E)=\bar{N}(E)+N_{fluc}(E).
\end{equation}

\noindent The smooth part $\bar{N}(E)$ is represented by the
generalized Weyl's law \cite{Baltes}

\begin{equation}\label{eq:19}
\bar{N}(E)\simeq\frac{A}{4\pi}E\mp\frac{L}{4\pi}\sqrt{E}+C
\end{equation}

\noindent
where minus and plus sign correspond to the Dirichlet and
the Neumann boundary condition, respectively. $A$ is a area of
billiard, $L$ is a length of the perimeter, and $C$ is a correction
constant for the curvature and corners given by

\begin{equation}\label{eq:20}
C=\frac{1}{12\pi}\int \kappa(s)\,ds +
\frac{1}{24}\sum_{i}\left(\frac{\pi}{\alpha_{i}}-\frac{\alpha_{i}}{\pi}\right)
\end{equation}

\noindent
with local curvature $\kappa(s)$ and $i_{th}$ corner angle $\alpha_{i}$.

The so-called $\delta_{n}$ quantity, which is equivalent to
$N_{fluc}(E_{n})$, is a good measure for the completeness of
obtained results

\begin{equation}\label{eq:21}
\delta_{n}=n-\frac{1}{2}-\bar{N}(E_{n})
\end{equation}

\noindent where $N(E_{n}):=n-\frac{1}{2}$. For complete eigenvalue
spectra, it has been well known that $\delta_{n}$ fluctuates around
zero. In Fig. 2(c), $\delta_{n}$ is drawn up to $1500$ level for the
$C_1$ continuity quartic basis. It certainly fluctuates around zero
and shows that the obtained eigenvalues are complete up to $1500$
level. Figure 2(c) also proves that the above criterion for the
relative error is acceptable.

\subsection{The Robnik billiard}

In this section, we consider the billiard which has been introduced
by Robnik \cite{Robnik1}. The boundary of the Robnik billiard is
defined by a quadratic conformal mapping from the unit circle

\begin{equation}\label{eq:22}
\omega=x+iy=z+\lambda z^{2}
\end{equation}

\noindent where $z=e^{i\phi}$, $\phi\in[0,2\pi]$, and
$\lambda\in[0,\frac{1}{2}]$. With increasing $\lambda$ from zero,
the boundary is continuously deformed from the unit circle. At
$\lambda=\frac{1}{2}$, the mapping of Eq. (22) is no longer
conformal and the billiard has a cusp. Such limit case of the Robnik
billiard is also called the cardioid billiard. It has a symmetry
line at $y=0$ and the desymmetrized billiard is twofold; odd and
even symmetry satisfying the Dirichlet and the Neumann boundary
condition at the symmetry line, respectively. In Fig. 3, we show the
cardioid billiard and its desymmetrized version.

\begin{figure}[b]
\includegraphics{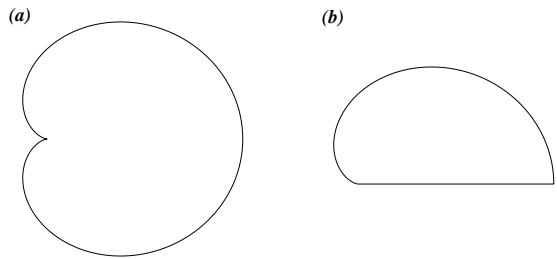}
\caption{The cardioid billiard, i.e., the limit case of the Robnik
billiard; (a) full and (b) desymmetrized version.}
\label{fig:fig3.eps}
\end{figure}

\begin{figure*}[t]
\includegraphics{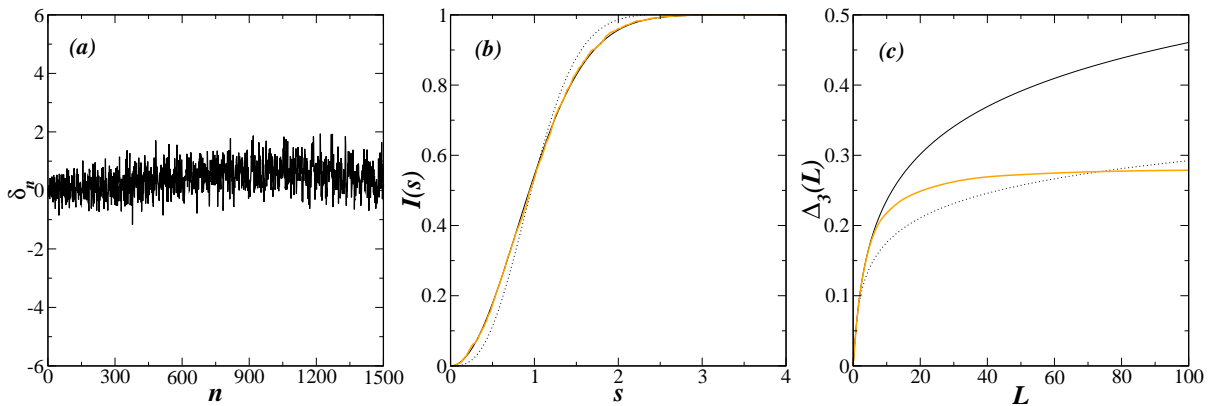}
\caption{(Color online) For the odd type of desymmetrized cardioid
billiard; (a) The $\delta_{n}$ up to $1500$ level. (b) The
cumulative level spacing distributions of the GOE, GUE, and the
considered billiard are drawn by a black full, black dotted, and
orange full line, respectively. (c) The spectral rigidities of the
GOE, GUE, and the considered billiard are drawn by a black full,
black dotted, and orange full line, respectively.}
\label{fig:fig4.eps}
\end{figure*}

It has been proven that the cardioid billiard is ergodic, mixing,
and a K system, i.e., fully chaotic system \cite{Markarian} and its
spectral statistics have been minutely studied in Ref.
\cite{Backer1}. We regard the odd type of desymmetrized cardioid
billiard as a model of convex geometric chaotic system and test the
FEM based on the $C_1$ continuity quartic interpolation. By applying
the method, we calculate $21901$ eigenvalues of the odd symmetric
case. Among those we obtain $1500$ accurate eigenvalue spectra. As
shown in Fig. 4(a), $\delta_{n}$ fluctuates around zero in this
range.

We investigate two spectral statistics, that is, the
nearest-neighbor level spacing distribution and the spectral
rigidity for $1500$ accurate eigenvalues. The nearest-neighbor level
spacing distribution $P(s)\,ds$ is the probability of finding a
consecutive pair of eigenstates for which the difference between
their eigenvalues lies in the interval $[s,s+ds]$. It measures the
short range correlation of the eigenvalue spectra. Instead of
$P(s)$, we consider the cumulative level spacing distribution

\begin{equation}\label{eq:23}
I(s)=\int_{0} ^{s} P(s^{'})\,ds^{'}
\end{equation}

\noindent to keep out of the binning problem about $P(s)$. The
spectral rigidity $\Delta_{3}(L)$ is the mean square deviation of
the spectral staircase function from the best fitting straight line
over a length $L$, namely

\begin{equation}\label{eq:24}
\Delta_{3}(L)=\Big\langle \min_{(a,b)} \frac{1}{L} \int_{-L/2} ^{L/2} d\epsilon\,
\big\{N(E+\epsilon)-a-b\epsilon\big\}^{2} \Big\rangle_E.
\end{equation}

\noindent It was firstly introduced by Dyson and Mehta \cite{Dyson}
to describe statistics of the energy levels of many particle systems
such as nuclei. It measures the long range correlation of the
eigenvalue spectra. Through studies about two mentioned spectral
statistics, we rescale the eigenvalue spectra $\{E_{i}\}$ into
$\{E_{i} ^{'}\}$ where $E_{i} ^{'}=\bar{N}(E_{i})$ and we omit the
prime. After the rescaling, the eigenvalue spectra have a mean level
spacing of unity and each billiard's own characteristic is contained
on the fluctuating part $N_{fluc}(E)$.

It has been widely accepted \cite{Bohigas} that the spectral
statistics of classically fully chaotic systems can be well
described by the universal laws of random matrix theory (RMT)
\cite{Mehta}. From the RMT prediction, the spectral statistics are
given by the distribution of the Gaussian orthogonal ensemble (GOE)
and the Gaussian unitary ensemble (GUE) for systems with and without
time reversal symmetry, respectively. Note that the time reversal
invariant systems possessing specific geometric properties can show
the GUE-like statistics. Concerned discussions will be addressed in
next section. From Berry's semiclassical analysis for spectral
rigidity \cite{Berry1}, it has been also known that the universality
region where the spectral statistics follow the universal RMT
prediction is finite. On paraphrasing, for fully chaotic systems,
the spectral rigidity $\Delta_{3}(L)$ shows a universal logarithmic
increase following the prediction of RMT in the interval $1\lesssim
L < L_{max}$. For the case of GOE, the coefficient of logarithm is
twice that of GUE. Then, in the range $L
> L_{max}$, $\Delta_{3}(L)$ reaches a non-universal saturation value
determined by short periodic orbits of corresponding classical
billiard. $L_{max}$ is called the outer energy scale and depends on
the period of shortest periodic orbit and the mean level density.

Since the cardioid billiard has the property of time reversal
invariance, it is expected that the cumulative level spacing
distribution $I(s)$ follows that of GOE and the spectral rigidity
$\Delta_{3}(L)$ is well described by the GOE prediction within the
universality regime. In Figs. 4(b) and 4(c), we show the results of
$I(s)$ and $\Delta_{3}(L)$ for the odd type of desymmetrized
cardioid billiard and compare with those of the GOE and GUE (for
numerical calculation of $I(s)$ and $\Delta_{3}(L)$ for the RMT
predictions, see the Ref. \cite{Backer1}). As expected, the results
show that the spectral statistics are in good agreement with the GOE
predictions and the spectral rigidity saturates beyond the
universality regime, which is restricted to small correlation length
L.

Note that we also calculate eigenvalue spectra for the even type of
desymmetrized cardioid billiard by applying the Neumann boundary
condition at the symmetry line. It can be easily achieved by the
$C_1$ continuity property of the applied shape function. We obtain
the same $1500$ accurate eigenvalues as the odd symmetric case. The
results of spectral statistics for the even symmetric case are well
described by the GOE expectation as qualitatively equivalent to the
odd symmetric case of Figs. 4(b) and 4(c). We would not present
these results in figure.

\subsection{The spiral-shaped billiard}

In this section we consider the spiral-shaped billiard whose
boundary $\partial\Omega$ is given by

\begin{equation}\label{eq:25}
r(\phi)=R\big(1+\epsilon\frac{\phi}{2\pi}\big)
\end{equation}

\noindent in polar coordinates $(r,\phi)$. $R$ is the radius of
spiral at $\phi=0$ and $\epsilon$ is the deformation parameter
determining relative size of the notch. The spiral-shaped billiard
is fully chaotic, that is, there is no stable island at all due to
its peculiar asymmetric property. We fix $R=1$ and consider two
cases of deformation parameter, namely, weakly deformed case at
$\epsilon=0.1$ drawn in Fig. 5(a) and strongly deformed case at
$\epsilon=0.3$ depicted in Fig. 5(b). We take the spiral-shaped
billiard as a model of non-convex geometric chaotic system and test
the numerical procedure presented in Sec. II.

\begin{figure}[b]
\includegraphics{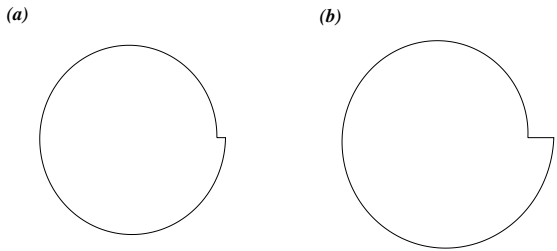}
\caption{The spiral-shaped billiards are shown on a same scale; (a)
weakly deformed case for $\epsilon=0.1$ and (b) strongly deformed
case for $\epsilon=0.3$.} \label{fig:fig5.eps}
\end{figure}

The spiral-shaped microcavity laser has been firstly introduced by
Chern {\it et al.} for obtaining unidirectional emission
\cite{Chern}. Afterwards, Lee {\it et al.} have found remarkable
resonance patterns of the spiral-shaped dielectric microcavity
exhibiting strong localizations on a {\it simple} geometric shape
\cite{Lee}. It looks like a clear counter-example of the
conventional scar-theory in which the localized intensity patterns
are appeared only on the corresponding classical unstable periodic
orbits \cite{Heller}, since the spiral has no simple-shaped periodic
orbit, that is, all periodic orbits must bounce the notch more than
once \cite{Lee1}. Recently, Lee {\it et al.} have shown that above
strongly localized resonance patterns can be approximated by linear
combinations of nearly degenerated resonance modes of the circular
cavity without any support from the classical periodic orbits
\cite{LeeJ}. Such recent research interests on the spiral-shaped
microcavity also motivate the studies on spectral statistics of the
spiral-shaped billiard.

First we consider the weakly deformed spiral-shaped billiard at
$\epsilon=0.1$. We obtain 1500 accurate eigenvalue spectra out of
$21955$ results calculated from the $C_1$ continuity quartic basis
FEM. The $\delta_{n}$ fluctuates around zero in this range as shown
in Fig. 6(a). In Figs. 6(b) and 6(c), we present the results of
$I(s)$ and $\Delta_{3}(L)$. We expect that the spectral statistics
are described by the GOE prediction due to the time reversal
invariance as the case of cardioid billiard. However the results
follow the GUE prediction rather than the GOE in Figs. 6(b) and
6(c).

For the strongly deformed spiral-shaped billiard at $\epsilon=0.3$,
we attain the same $1500$ accurate eigenvalues among $21934$
calculated data (see the results of $\delta_{n}$ in Fig. 7(a)). In
contrast to the weakly deformed case, the results of $I(s)$ and
$\Delta_{3}(L)$ are well described by the GOE expectation as one can
show in Figs. 7(b) and 7(c). In result, different degrees of
deformation causes quite different spectral statistics and the
unexpected GUE-like statistics are observed.

In the literatures, there have been several reports that study the
spectral statistics exhibiting unexpected GUE-like behavior in time
reversal invariant systems, for examples, the system with certain
point symmetry \cite{Leyvraz,Dietz} and the so-called Monza billiard
possessing the property of unidirectional motion \cite{Veble}. For
these systems it has been known that the GUE-like spectral
statistics have their origin in the degenerated eigenstates. However
the GUE-like statistics of the weakly deformed spiral-shaped
billiard at $\epsilon=0.1$ cannot be explained by this reason, since
the system has no degenerated eigenstates. Note that
reasonable accounts for the unexpected spectral statistics of the
weakly deformed spiral-shaped billiard are not feasible at present.
But we would anticipate that all of above GUE-like spectral
statistics can be understood in an unified principle.

\section{conclusion}

We present an efficient finite element method for calculating
eigenvalues and eigenfunctions of quantum billiard systems. The
$C_1$ continuity quartic interpolation basis is considered. We show
that the method provides accurate set of eigenvalues exceeding a
thousand levels for any shape of quantum billiards on a personal
computer. Comparison with the well-known $C_0$ continuity quadratic
basis FEM proves the efficiency of the applied method. The spectral
statistics of the Robnik and the spiral-shaped billiards are studied
and the unexpected GUE-like behaviors are observed.

Note that we do not make use of the sparsity of matrices. We would
expect that the generalized eigenvalue solving routine, which is
optimized to sparse matrix, enhances the efficiency of presented
FEM.

\begin{figure*}[t]
\includegraphics{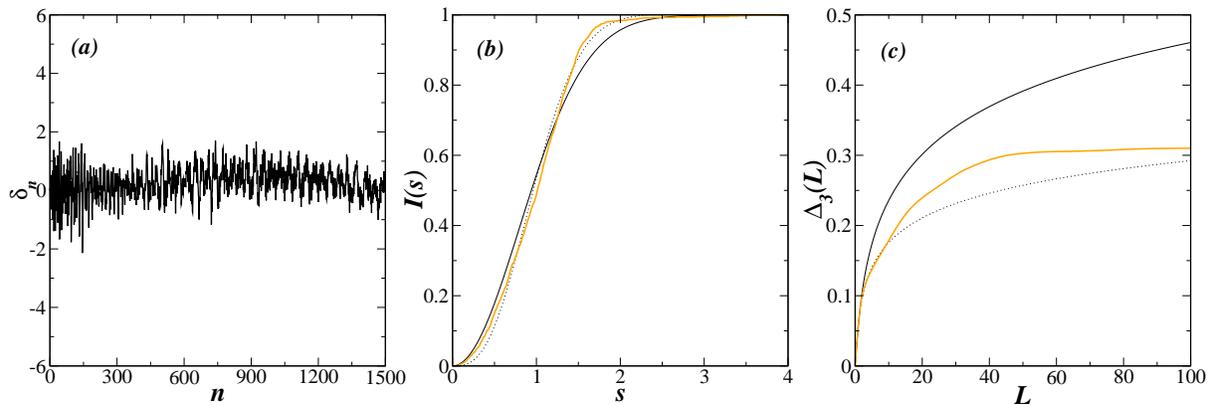}
\caption{(Color online) For the weakly deformed spiral-shaped
billiard at $\epsilon=0.1$; (a) The $\delta_{n}$ up to $1500$ level.
(b) The cumulative level spacing distributions of the GOE, GUE, and
the considered billiard are drawn by a black full, black dotted, and
orange full line, respectively. (c) The spectral rigidities of the
GOE, GUE, and the considered billiard are drawn by a black full,
black dotted, and orange full line, respectively.}
\label{fig:fig6.eps}
\end{figure*}

\begin{figure*}[t]
\includegraphics{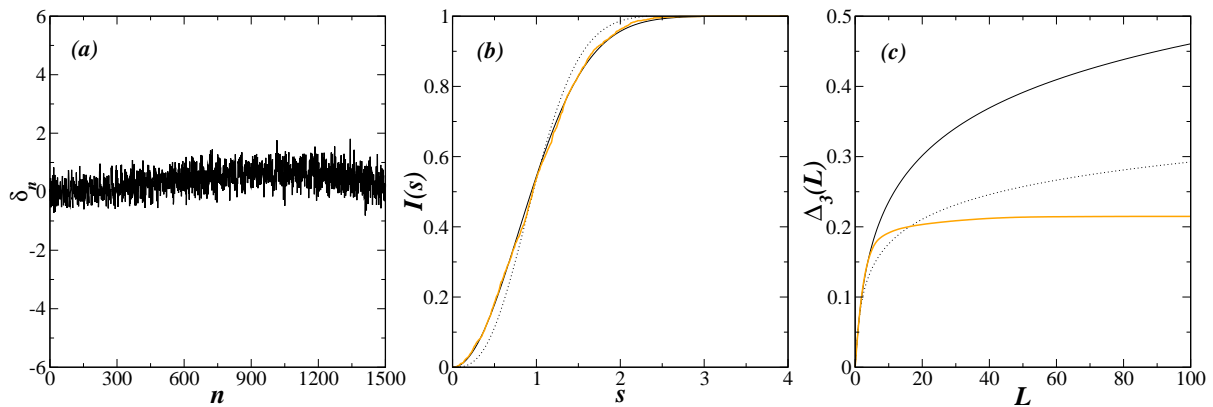}
\caption{(Color online)
For the strongly deformed spiral-shaped billiard at $\epsilon=0.3$;
(a), (b), and (c) contain the same quantities of Fig. 6.}
\label{fig:fig7.eps}
\end{figure*}

\begin{acknowledgments}
This study was supported by Acceleration Research
(Center for Quantum Chaos Applications) of MEST/KOSEF.
\end{acknowledgments}

\end{document}